# A bidirectional dual-comb ring laser for simple and robust dual-comb spectroscopy

Takuro Ideguchi[1,*], Tasuku Nakamura[2], Yohei Kobayashi[3] & Keisuke Goda[2,4,5,*]

[1]Research Centre for Spectrochemistry, University of Tokyo, Tokyo 113-0033, Japan
[2]Department of Chemistry, University of Tokyo, Tokyo 113-0033, Japan
[3]The Institute for Solid State Physics, University of Tokyo, Kashiwa 277-8581, Japan
[4]Department of Electrical Engineering, University of California, Los Angeles, CA 90095, USA
[5]Japan Science and Technology Agency, Tokyo 102-0076, Japan
[*] Corresponding authors: ideguchi@chem.s.u-tokyo.ac.jp, goda@chem.s.u-tokyo.ac.jp

**Fourier-transform spectroscopy[1] is an indispensable tool for analyzing chemical samples in scientific research as well as chemical and pharmaceutical industries[2-7]. Recently, its measurement speed, sensitivity, and precision have been shown to be significantly enhanced by using dual frequency combs[8-19]. However, wide acceptance of this technique is hindered by its requirement for two frequency combs and active stabilization of the combs. Here we overcome this predicament with a Kerr-lens mode-locked bidirectional ring laser that generates two frequency combs with slightly different pulse repetition rates and a tunable yet highly stable rate difference. This peculiar lasing principle builds on a slight difference in optical cavity length between two counter-propagating lasing modes due to Kerr lensing. Since these combs are produced by the one and same laser cavity, their relative coherence stays passively stable without the need for active stabilization. To show its utility, we demonstrate broadband dual-comb spectroscopy with the single laser.**

Since it was commercialized more than 40 years ago, the Fourier-transform spectrometer based on Michelson-based Fourier-transform spectroscopy[1], most notably known as a form of Fourier-transform infrared spectroscopy (FTIR), has long been used as a user-friendly, robust, and reliable instrument for analyzing and identifying chemical compounds in diverse fields such as molecular precision spectroscopy[2], organic synthesis, chemical catalysis[3], biological analysis[4,5], environmental monitoring, and clinical pathology[6,7]. Over the last decade, it has been proven that its measurement speed, sensitivity, and precision can significantly be enhanced[8-19] by using laser frequency combs[20]. This technique known as dual-comb spectroscopy is based on the principle (analogous to the principle of a sampling oscilloscope) that the probe light composed of two frequency combs with slightly different pulse repetition rates works as a mechanical-scan-less temporal interferometer due to their linearly increasing relative time delay between the pulse pairs. In the frequency domain, it can be understood as a multi-heterodyne detection of the pairs of neighboring comb lines that have slightly different line spacings by which the optical frequency combs are down-converted to a single radio-frequency comb. This scan-less interferometer has been demonstrated to outperform the traditional mechanically-driven Michelson-based Fourier-transform spectrometer by several orders of magnitude in terms of measurement time, sensitivity, and spectral resolution (hence precision).

Unfortunately, despite its excellent capabilities, dual-comb spectroscopy has not been adopted beyond research laboratories. This is due to its costly requirement for two laser frequency combs and comb stabilization systems such as f-2f interferometers or external reference cavities with feedback control circuits. Since mode-locked lasers (free-running combs) have independent fluctuations in pulse repetition rate and carrier-envelope phase, their dual-comb interferogram is significantly distorted, spoiling its Fourier transform (spectrum) with chromatic artifacts. For this reason, a frequency stabilization system is typically required to tightly phase-lock either frequency comb against the other comb and hence to satisfy the requirement for the high relative coherence between the combs in order to fully benefit from the superior performance of dual-comb spectroscopy. However, such high-precision comb-stabilization techniques are only available at optical metrology laboratories with considerable expertise and infrastructure. Consequently, the high cost and complexity of the dual-comb spectrometer betray the virtue of the simple and handy Fourier-transform spectrometer for which it has been widely used by diverse researchers for the last 40 years.

In this Letter, to overcome this difficulty, we propose and demonstrate a surprisingly simple yet robust method for dual-comb spectroscopy based on a bidirectional ring laser. The laser is composed of a free-running Kerr-lens mode-locked laser cavity that produces two laser frequency combs with slightly different pulse repetition rates and a tunable yet highly stable rate difference. This lasing principle, which we refer to as direction-dependent lasing due to asymmetrical Kerr lensing, is based on our revisit to and further development of the effect that there exists a slight difference in optical cavity length between the two counter-propagating lasing modes (Supplementary Figure 1) – an effect that is contrary to the photonics community's traditional understanding that the two laser outputs should have the same pulse repetition rate as they share the same optical cavity. Since these combs are produced by the one and same laser cavity, their relative coherence stays passively stable. As a result, it enables highly stable broadband dual-comb spectroscopy without the need for any mechanical moving parts and complicated comb stabilization systems with feedback control circuits. As a proof-of-principle demonstration, we show broadband dual-comb spectroscopy with the single free-running laser. This simple and robust dual-comb laser is expected to be highly effective for in situ spectroscopy in a diverse range of fields including food science, environmental science, forensic science, pharmaceutical science, planetary science, medicine, and manufacturing.

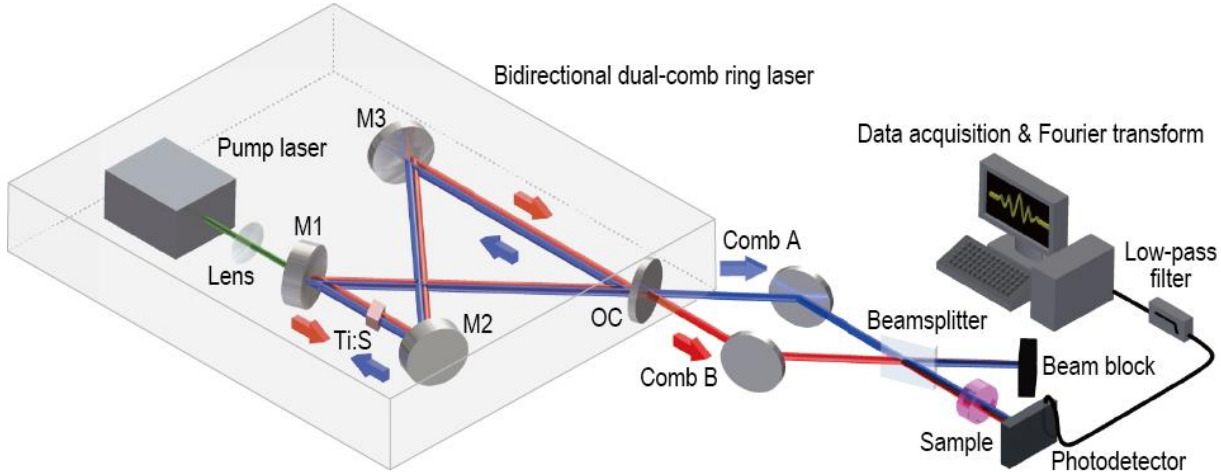

**Figure 1 | Schematic of the bidirectional dual-comb ring laser and dual-comb spectroscopy system.** The laser is based on a four-mirror bow-tie ring cavity with a Ti:Sapphire (Ti:S) crystal pumped by a diode-pumped-solid-state laser at 532 nm. All the cavity mirrors including two concave mirrors (M1, M2), a convex mirror (M3), and an output coupler (OC) are dielectric-coated to provide negative group-delay dispersion (GDD) for intra-cavity pulse compression. The ring cavity generates ultrashort pulse trains by means of soft-aperture Kerr-lens mode-locking in either uni-directional or bi-directional lasing mode. In the case of bidirectional mode-locking, the repetition rates of the two outputs can be made identical or slightly different, depending on the cavity alignment. The two outputs of the laser (Comb A, Comb B), spatially combined by a beamsplitter, pass through the sample and are incident onto the photodetector. The low-pass-filtered photodetector signal is digitized and Fourier-transformed for dual-comb spectroscopy.

As schematically shown in Figure 1, the bidirectional dual-comb ring laser is based on a four-mirror bow-tie ring cavity with a Ti:Sapphire crystal pumped by a diode-pumped-solid-state laser at 532 nm with an average power of approximately 9 W (a similar design can be found in Ref. 21). All the cavity mirrors including two concave mirrors (M1, M2), a convex mirror (M3), and an output coupler (OC) are dielectric-coated to provide negative group-delay dispersion (GDD) for intra-cavity pulse compression. The ring cavity generates ultrashort pulse trains by means of soft-aperture Kerr-lens mode-locking[22] in either uni-directional or bi-directional lasing mode[23]. In the case of bidirectional mode-locking, the repetition rates of the two outputs can be made identical or slightly different, depending on the cavity alignment. The origin of the direction-dependent lasing effect, namely the difference in repetition rate between the two pulse trains, can be explained by a slight difference in round-trip optical path length between the two counter-propagating lasing modes due to the direction-dependent difference in nonlinear refractive index within the laser crystal, which may arise from the difference in beam diameter or overlap with the pump beam in the crystal between the counter-propagation pulses (Supplementary Figure 1).

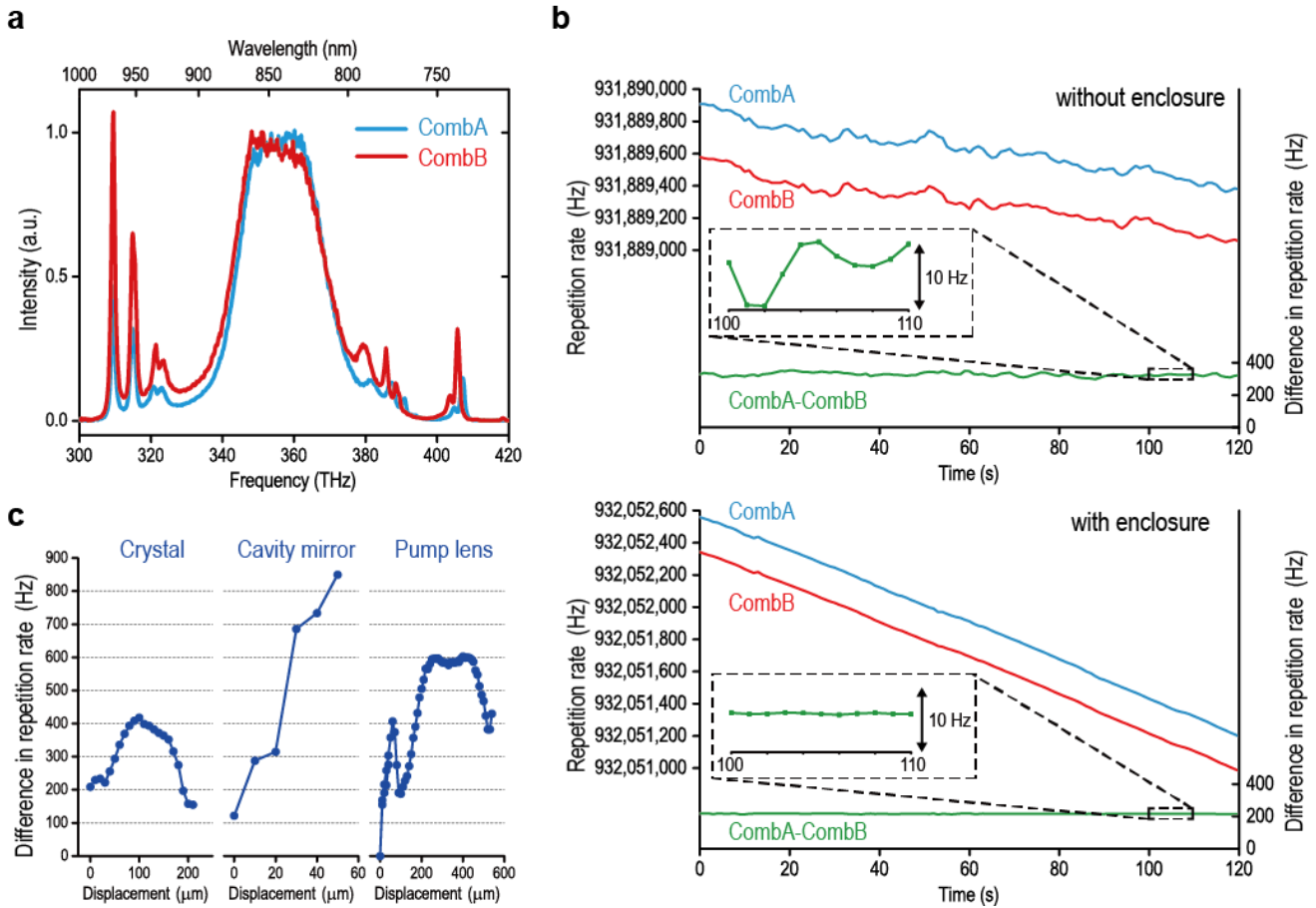

**Figure 2 | Basic performance of the bidirectional dual-comb ring laser. a**, Spectra of the two laser outputs (shown in blue and red) measured by a grating-based spectrometer. **b**, Temporal behavior of the pulse repetition rates of the two combs (shown in blue and red) and the difference between them (shown in green). The insets show the zoomed difference in pulse repetition rate over 10 seconds. **c**, Tunability of the difference in pulse repetition rate between the two laser outputs as a function of the displacement of the Ti:Sapphire crystal, the cavity mirror M2, and the focusing lens for the pump laser with respect to the initial positions where the laser has been mode-locked. The positive directions of the displacements (x axis) in the figure are defined as the forward, forward, and backward directions of the pump laser's propagation for the crystal, mirror, and lens, respectively.

The spectra of the two outputs of the bidirectional dual-comb ring laser are shown in Figure 2a. They have nearly identical spectral shapes with a center frequency of 358 THz and a FWHM bandwidth of 26 THz (corresponding to 61 nm in wavelength). The slight difference in shape between the two spectra indicates non-identical lasing between the counter-propagating modes in the cavity. The output pulses are slightly chirped, but can be compressed down to 12 fs by an external dispersion compensator. The two combs have an intensity profile of $TEM_{00}$ and an average optical power of 340 mW and 310 mW, respectively. The broad bandwidth (26 THz) of the laser is essential for broadband dual-comb spectroscopy and superior to previously reported single-oscillator-based mode-locked dual-comb lasers based on a semiconductor disk laser[24] or a monolithic cavity laser[25] that only provide narrow bandwidths (less than 300 GHz). With simple modifications, the dual-comb laser can be configured to produce the spectrally broadest (up to 175 THz) and temporally shortest (down to 5.4 fs) pulses directly from the oscillator[26] with the lowest noise level[27].

The pulse repetition rate of each laser output and the difference in pulse repetition rate between them are shown in Figure 2b. While each repetition rate fluctuates around 932 MHz due to environmental disturbances (e.g., air flow, ambient temperature fluctuations, and mechanical vibrations), the difference between them remains highly stable at a quasi-constant value of 325 Hz with a standard deviation of 3.4 Hz over minutes. This common-mode noise of the two repetition rates is a significant advantage for dual-comb spectroscopy in which the difference in pulse repetition rate between the two combs must be kept constant during the measurement of interferograms. The repetition rate difference can further be made more stable by, for example, enclosing the laser cavity with a simple housing made of aluminum (Figure 2b). The figure indicates a drastic improvement in the stability, resulting in a standard deviation of 0.1 Hz around a repetition rate difference of 216 Hz. We anticipate that a hermetically sealed cavity, commonly employed for commercial products, can further stabilize the repetition rate difference against air flow and ambient temperature fluctuations.

The difference in pulse repetition rate between the two laser outputs can be turned over ~1 kHz by multiple techniques. As shown in Figure 2c, it can be varied by displacing either the Ti:Sapphire crystal, the cavity mirror (M2), or the focusing lens for the pump laser with the laser pulses mode-locked during the displacement. While the mirror displacement has a high frequency tunability of 15 Hz/µm over a large frequency range, the crystal and lens displacements have a relatively low tunability of 3 Hz/µm (evaluated in the quasi-linear region) and work in a narrower frequency range. These features can be taken advantage of for high-speed modulation of the repetition rate difference using the mirror displacement[9] and for precise tuning of the difference using the crystal and lens displacements.

The relative temporal coherence between the two pulse trains is directly evaluated by measuring the linewidth of the radio-frequency beat note between a pair of comb lines from the two combs (See Methods for details). As shown in Figure 3a, the beat note has a FWHM linewidth of 13 kHz in a measurement duration of 1 ms, indicating that the interferogram measured by the dual-comb spectrometer can maintain coherence for 77 µs. In addition, as shown in Figure 3b, the beat note is kept stable within 200 kHz over 60 minutes, indicating high long-term stability in the relative coherence between the two pulse trains. This long-term stability firmly supports continuous measurements for a long duration of time without suffering from a detrimental aliasing effect that would otherwise occur in the case of unstable coherence.

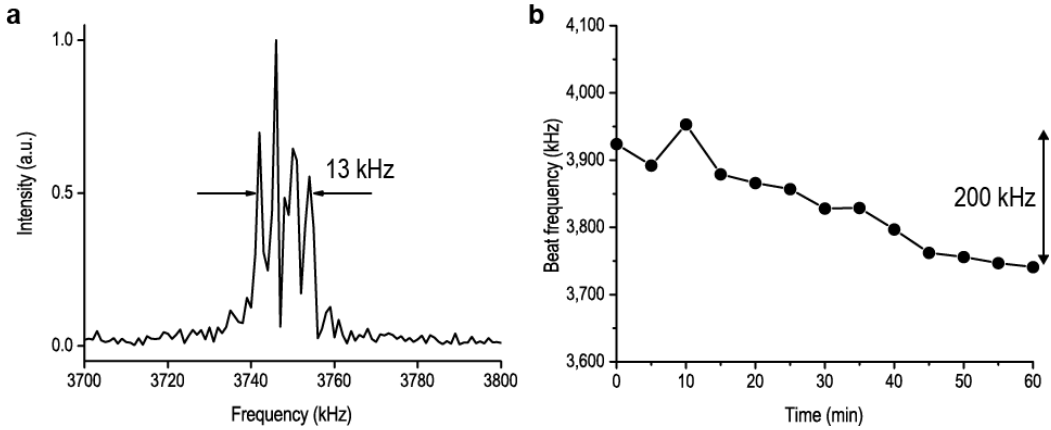

**Figure 3 | Relative coherence between the two combs. a**, Beat note between a pair of comb lines from the two combs measured in 1 ms, showing high relative coherence between them. **b**, Frequency of the beat note sampled at every 5 minutes over 60 minutes, showing high long-term stability in the relative coherence between the two combs.

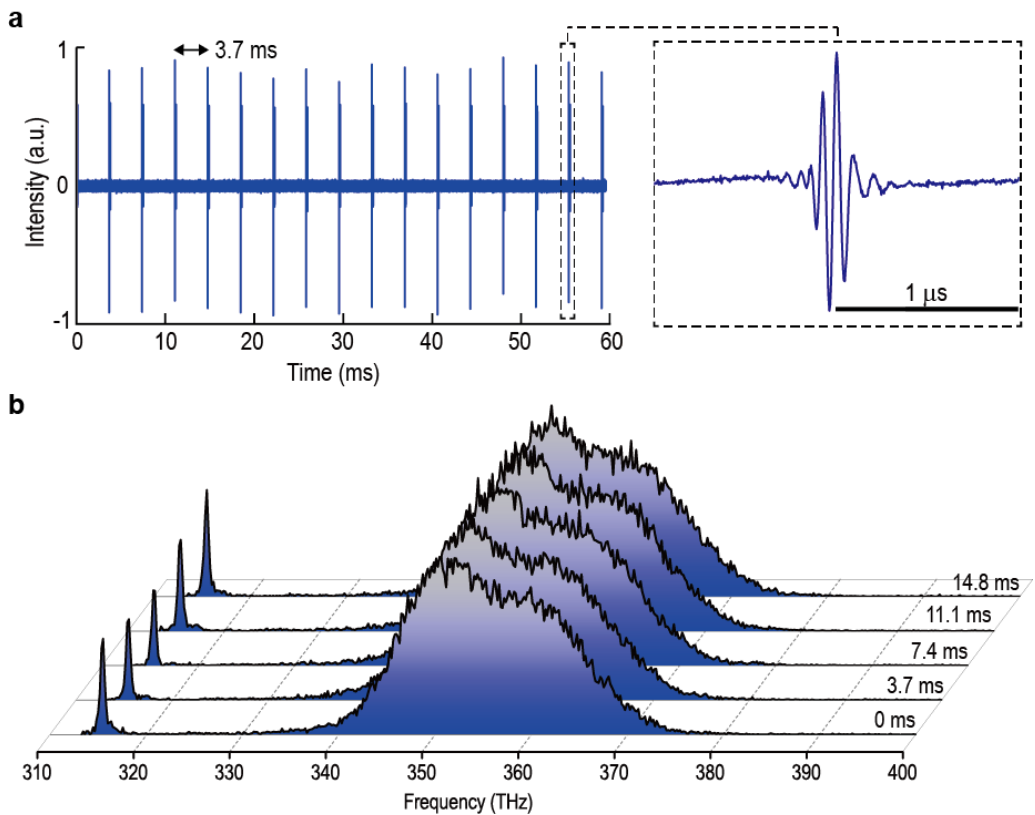

**Figure 4 | Dual-comb spectroscopy with the bidirectional dual-comb ring laser. a**, Continuously measured signal that includes multiple interferograms. The interval between the consecutive interferograms is 3.7 ms. The inset shows one of the interferograms. **b**, Consecutive broadband dual-comb spectra that are obtained by Fourier-transforming each interferogram over 20 μs around each burst point.

Finally, we performed dual-comb spectroscopy with the bidirectional dual-comb ring laser. Figure 4a shows a sequence of dual-comb interferograms acquired by detecting two spatially overlapped frequency combs with a repetition frequency difference of 270 Hz (without a sample). Fourier-transforming each interferogram over 20 μs around each burst point produces a broadband dual-comb spectrum that can repetitively be acquired every 3.7 ms (Figure 4b). We then conducted dual-comb absorption spectroscopy of a Nd:YVO$_4$ crystal with a spectral resolution of 93 GHz and an acquisition duration of 67 μs over a spectral range of 18 THz (corresponding to 40 nm) centered at 367 THz (corresponding to 817 nm) (Figure 5). The dual-comb spectrum clearly shows the absorption lines of Stark-split sub-levels from the ground level ($^4I_{9/2}$) to the $^2H_{9/2}$ and $^4F_{5/2}$ bands of neodymium ion (Nd$^{3+}$) doped into yttrium orthovanadate (YVO$_4$). The spectrum is in good agreement with the spectrum of the same sample obtained by a conventional grating-based optical spectrum analyzer (OSA).

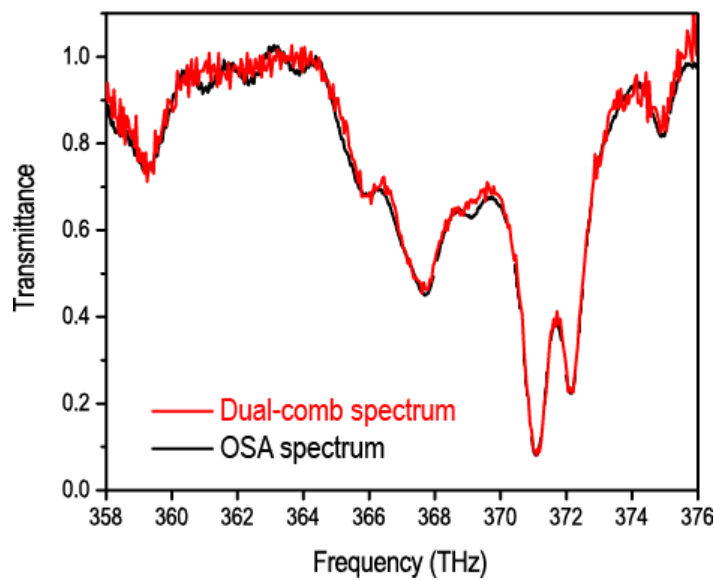

**Figure 5 | Dual-comb absorption spectroscopy of a Nd:YVO$_4$ crystal with the bidirectional dual-comb ring laser.** The dual-comb absorption spectrum of a Nd:YVO$_4$ crystal was obtained in transmission mode (shown in red) with a spectral resolution of 93 GHz and an acquisition duration of 67 μs over a spectral range of 18 THz (corresponding to 40 nm) centered at 367 THz (corresponding to 817 nm). The spectrum agrees well with the spectrum of the same sample obtained by a conventional grating-based optical spectrum analyzer (black).

The capabilities of the dual-comb spectrometer with the dual-comb laser can further be enhanced by incorporating a few techniques into it or extending it beyond what has been demonstrated in this Letter. First, phase-error-correction techniques such as real-time adaptive sampling[13] and post computational re-sampling[10] can easily be implemented on the system in order to achieve higher spectral resolution. Second, while in our proof-of-principle demonstration, we have performed spectroscopy in the near-infrared region

(750 - 950 nm), the dual-comb laser can be extended to other spectral regions such as XUV, UV, visible, mid-infrared, and THz via nonlinear frequency conversion or spectral broadening. Third, the dual-comb laser can also be used for other dual-comb applications than dual-comb spectroscopy, such as asynchronous optical sampling[28], absolute long-range distance measurements[29], and optical coherence tomography[30]. Finally, by virtue of its ultrashort pulses, the dual-comb laser can directly be employed for nonlinear spectroscopy such as coherent Raman scattering spectroscopy[17] and two-photon excitation spectroscopy[19].

## Methods
### Design of the dual-comb laser
The Ti:Sapphire mode-locked laser is based on a bow-tie ring cavity that consists of four mirrors including two concave mirrors (M1, M2) with a radius of 30 mm, a convex mirror (M3) with a radius of 1,000 mm, and a flat output coupler (OC). All of the mirrors are dielectric-coated to achieve high reflectivity in a broad spectral range: >99.9 % (620-1050 nm) for the concave mirrors, >99.8 % (670-1000 nm) for the convex mirror, and 99 % (700-900 nm) for the output coupler. The mirror coatings are also designed to add negative dispersion with -50 $fs^2$, -60 $fs^2$, and <20 $fs^2$ for the concave mirrors, the convex mirror, and the output coupler, respectively. The Ti:Sapphire crystal with a thickness of 2.3 mm is placed at a point near the middle point between M1 and M2 at a Brewster angle against the beam path. The crystal is pumped by a CW DPSS green laser at 532 nm (Millenia eV, Spectra Physics) with an average power of 8-9 W focused by a plano-convex lens with a focal length of 40 mm.

### Evaluation of the dual-comb laser
The spectra of the laser outputs shown in Figure 2a were measured by a grating-based spectrometer with a resolution of 400 GHz (LSM-Mini, Ocean Photonics). The repetition rates shown in Figure 2b and Figure 2c were measured in a gate time duration of 1 s with two frequency counters (FCA3003, Tektronix) referenced to a 10 MHz rubidium frequency standard (FS725, Stanford Research Systems).

### Measurement of the beat note between a pair of comb lines from the two outputs of the dual-comb laser
For the beat note measurement, a CW external-cavity diode laser running at 790 nm (TLK-L780M, Thorlabs) was used as an intermediate. First, the beat note between the CW laser and one of the combs was generated by a fast photodetector, from which the beat note at the lowest frequency was extracted by a low-pass filter. The beat note between the CW laser and the other comb was also generated in the same way. Next, the beat notes were mixed by an electronic mixer, from which only the subtraction signal was extracted by filtering. Finally, the signal was digitized by a 14-bit data acquisition board (ATS9440, Alazartech) at a sampling rate of 20 MS/s. Fourier transformation of the sampled data yields the beat note shown in Figure 3a. To produce Figure 3b, the beat-note measurements were performed every 5 minutes.

**Dual-comb spectroscopy with the dual-comb laser**

For dual-comb spectroscopy, the two outputs were combined by a 50/50 beamsplitter into a dual-comb beam, which is incident onto an amplified high-speed Si photodetector. The detector signal was low-pass-filtered at a cutoff frequency of less than half of the repetition rate of the combs (<466 MHz) and digitized by a 14-bit data acquisition board. For the absorption spectroscopy measurement shown in Figure 5, a Nd:YVO$_4$ crystal was placed as a sample in between the beamsplitter and the photodetector. A short-pass filter with a cutoff wavelength of 850 nm was placed in front of the photodetector.

**Acknowledgements**

We are grateful to K. Yoshioka for the loan of optical components. This work was supported by the ImPACT Program of the Council for Science, Technology and Innovation (Cabinet Office, Government of Japan), Burroughs Wellcome Foundation, Sumitomo Foundation, and Konica Minolta Imaging Science Award.

**Author contributions**

T. I designed the work. T. I. and T. N. performed the experiments and data analysis. Y. K. designed and installed the laser. K. G. supervised the work. T. I. and K. G. wrote the manuscript. T. N. and Y. K. provided comments and suggestions for improvements in the manuscript.

# Supplementary Information

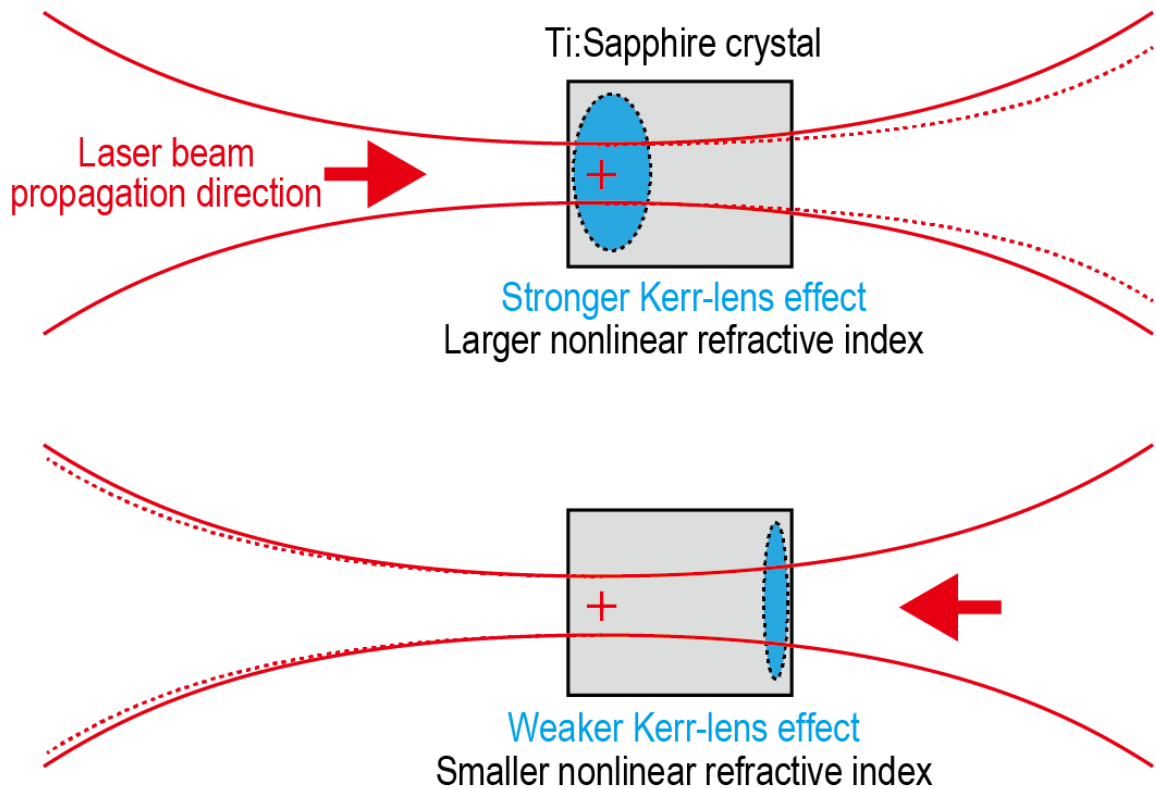

**Supplementary Figure 1 | Direction-dependent lasing.** The origin of the direction-dependent lasing effect, namely the difference in repetition rate between the two pulse trains, can be explained by a slight difference in optical path length between the two counter-propagating lasing modes due to the direction-dependent difference in nonlinear refractive index within the laser crystal. This may arise from the difference in beam diameter or overlap with the pump beam in the crystal between the counter-propagation pulses. The interplay between lasing and the Kerr-lensing effect leads to the generation of two pulse trains with slightly different repetition rates.